\documentclass[runningheads]{llncs}
\usepackage{graphicx}

\usepackage{tikz}
\usepackage{PRIMEarxiv}
\usepackage{amsmath,amssymb} 
\usepackage{color}
\usepackage{float}
\usepackage[colorlinks,linkcolor=black]{hyperref}

\usepackage[accsupp]{axessibility}  

\usepackage{graphicx}
\usepackage{subfig}

\begin{document}
\pagestyle{headings}
\mainmatter

\title{Content-Based Landmark Retrieval
Combining Global and Local Features
using Siamese Neural Networks} 



\author{
   Tianyi Hu \\
   Technische Universität Berlin \\
   Germany, Berlin \\
   \And
  Monika Kwiatkowski \\
  Computer Vision \& Remote Sensing \\
  Technische Universität Berlin \\
   Germany, Berlin \\
   \And
  Simon Matern \\
  Computer Vision \& Remote Sensing \\
  Technische Universität Berlin \\
   Germany, Berlin \\
     \And
  Olaf Hellwich \\
  Computer Vision \& Remote Sensing \\
  Technische Universität Berlin \\
   Germany, Berlin \\
}
\maketitle

\begin{abstract}

In this work, we present a method for landmark retrieval that utilizes global and local features. 
A Siamese network is used for global feature extraction and metric learning, which gives an initial ranking of the landmark search. 
We utilize the extracted feature maps from the Siamese architecture as local descriptors, the search results are then further refined using a cosine similarity between local descriptors. 
We conduct a deeper analysis of the Google Landmark Dataset, which is used for evaluation, and augment the dataset to handle various intra-class variances. 
Furthermore, we conduct several experiments to compare the effects of transfer learning and metric learning, as well as experiments using other local descriptors. 
We show that a re-ranking using local features can improve the search results. 
We believe that the proposed local feature extraction using cosine similarity is a simple approach that can be extended to many other retrieval tasks. 

\end{abstract}

\section{Introduction}
With the development of computer technology and the application of 5G networks, 
images have gradually become the main carrier of and dissemination method for information.
The Content-Based Image Retrieval (CBIR) system, based on the similarity of graphic image shape, edge pattern, and color scheme, enables users to retrieve massive amounts of graphical information efficiently and conveniently \cite{chen2021deep}. 
Landmark retrieval is a special kind of CBIR that focuses on landmark images. 
Many landmark retrieval datasets have low inter-class variance, e.g. different classes of churches are very similar to each other \cite{gavves2012visual}. 
Therefore, the separability between the categories is often more difficult than for other CBIR tasks. Additionally, the variance within the categories is relatively large due to changes in image perspective, varying time of image acquisition. Besides, some images were taken inside of a landmark building, while some were taken from the outside, which further increases intra-class variance. \\


Existing landmark retrieval methods use deep learning networks to extract global features of landmark images, compare the features between different image pairs to train the model, and then retrieve data from the database according to the trained model \cite{chen2021deep}. 
In addition, some methods use the local features of the image to perform feature point matching to retrieve images. 
The global features differ from the local features in several respects. 
Global features generally describe the overall content of the image, and common global features refer to color, textures and shape, etc. 
However, these features pose difficulties when retrieving the desired landmark in the case of occlusions or people in the retrieval task \cite{jegou2011aggregating} \cite{gordo2017end} \cite{radenovic2018fine}. 
In contrast, the local features of the image generally describe the edges, key points, or certain specific areas of the image. 
The local features produce good retrieval results for specific image retrieval tasks, such as when the query image is occluded or flipped. 
However, the local features of the image cannot reflect the full content of the image \cite{lowe2004distinctive} \cite{sanchez2013image} \cite{jegou2010aggregating}, therefore, it is better to combine the two for re-ranking in retrieval tasks \cite{cao2020unifying}. \\

These methods have a specific problem in the Landmark Dataset: 
the intra-class variance is large and the inter-class variance is low between similar type of landmarks, which makes it difficult to achieve a high retrieval accuracy using global features or local features alone \cite{cao2020unifying}. 
In order to solve this problem, in this paper, we build a \textit{Global-Local} approach. 
First, we apply transfer-learning to a pre-trained model EfficientNet-B0 \cite{tan2019efficientnet} to the Google Landmark Dataset v2.1 (GLD v2.1) and then use the metric learning method for training which gives the global feature extractor. 
Next, we use the patch retrieval approach to re-rank the top 100 images retrieved by the global feature based retrieval method to optimize the results. 
Through this method, the specific problem of low retrieval accuracy in the Landmark Dataset can be improved. 
Due to the limitations of computation time and model format, online evaluation is not suitable for the comparison, and all the evaluations are implemented offline. 
In the end, our search results reach 33.9\% on the private dataset and 32.84\% on the public dataset using mAP@100. 

Our contributions are summarized as follows: 
 \begin{itemize}
     \item \noindent We build a two stage network ($Global-Local$) to improve the accuracy of landmark retrieval, consisting of global and local feature extractors. 
     Global features are extracted using EfficientNet-B0 and then the EfficientNet-B0, and fine-tuned using transfer learning and metric learning. 
     Local features are extracted using the patch retrieval approach. 
     \item \noindent We combine the global and local features to implement the re-ranking in the GLD v2.1 data set. 
     \item \noindent Compared with the traditional neural network, the experimental results have significantly improved the accuracy of the landmark retrieval task. 
 \end{itemize}


\section{Related Work}
\subsubsection*{Image retrieval based on global features.}
The global features of an image describe its overall characteristics, including the shape, color, texture, etc. 
Shape-based retrieval methods include Fourier Descriptor \cite{kauppinen1995experimental}, Generic Fourier Descriptor \cite{zhang2002generic}, boundary direction histogram \cite{2004Shape}, etc. 
Color-based retrieval methods include Color Histogram \cite{swain1991color},
Color Constant Color Index method \cite{funt1995color}, Octree \cite{wan1998new}, etc. 
Texture-based retrieval methods include Local Binary Patterns (LBP) algorithm \cite{ojala2002multiresolution}, gray level co-occurrence matrix \cite{tuceryan1993texture}, etc. 
These global feature extraction methods use a single feature of an image for extraction. 
All the methods have rely on a specific global feature, and their accuracy is limited for large and complex datasets.
Recently, with the development of deep networks, high performance global feature extraction methods are all based on deep learning, such as
\cite{tolias2015particular} \cite{gordo2017end} \cite{radenovic2018fine}. 
The features extracted through deep learning are a high dimensional embedding, which contains more contextual information and enables a more robust image retrieval.
\subsubsection*{Image retrieval based on local features.}
The local features of an image mainly describe special areas or edge points. 
They ensure good retrieval results under circumstances where some areas of the query images are obscured. 
Early local feature extraction methods use SIFT \cite{lowe2004distinctive} or Speeded Up Robust Features (SURF)\cite{bay2008speeded} to extract feature descriptors in query and dataset images, calculating similarities sequentially. This method is robust but slow. 
In recent years, methods to extract local features of images based on deep-learning have been introduced, for example by Noh et al  \cite{noh2017large} and Mishchuk et al \cite{mishchuk2017working}. 
\subsubsection*{Image retrieval based on global features and local features.}
Image retrieval methods based on global and local features have become a trend in recent years, as the accuracy of image retrieval can be improved by combining the two types of features. 
Cao~\textit{et al.}\cite{cao2020unifying} employs ResNet as the backbone, chooses average pooling layer to extract features, and uses attention-based keypoints to extract local features of images. 
This approach fuses the two features in one network for end-to-end training to achieve the best accuracy. 

\section{Preliminary and Background}
\begin{figure}[htbp]
\centering
\includegraphics[height=0.4\linewidth]{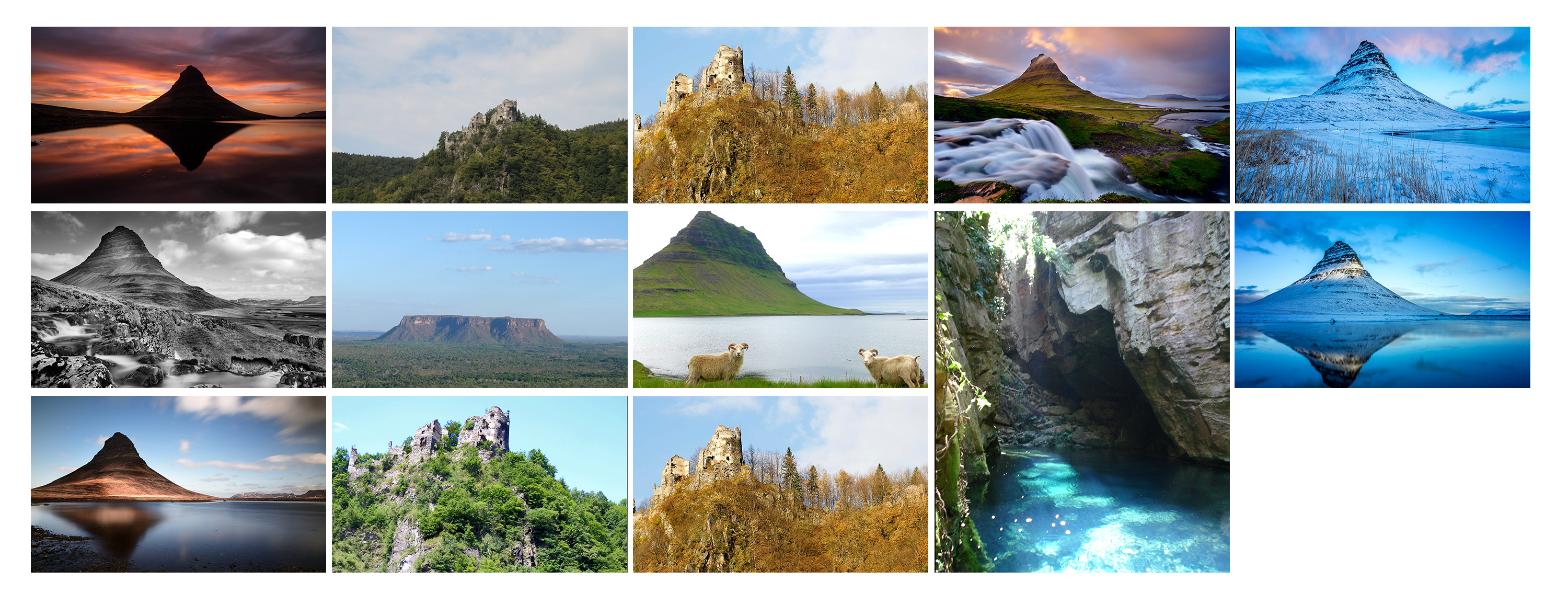}
\caption{Sample image from GLD v2.1 data set.}
\label{Sample image from GLD-v2.1 data set}
\end{figure}

In this section, we investigate the GLD v2.1 data set and conduct a descriptive data analysis on it. 
Our goal is to examine the character of the data set, which is beneficial for designing the architecture of the neural network and designing a data augmentation pipeline. 
The original dataset has over 5 million training images and 200,000 categories. 
However, the original training dataset contains a large number of noisy images, which will greatly affect our retrieval results. 
Hence, the team smlyaka \cite{yokoo2020two} uses an automated data cleaning method to eliminate the noise, which results in the cleaned GLD v2.1 data set consisting of 1,580,470 images and 81,313 categories\footnote{This dataset is obtained from \url{https://github.com/cvdfoundation/google-landmark}.}. 
The rest of the original training data set consists of 2,553,444 noisy images, most of which are not landmark images. 
For better retrieval we use team smlyaka's cleaned version. 
Landmark images of natural scenes are depicted in Fig.~\ref{Sample image from GLD-v2.1 data set}. 

\begin{figure}[htbp]
\centering
\includegraphics[height=0.5\linewidth]{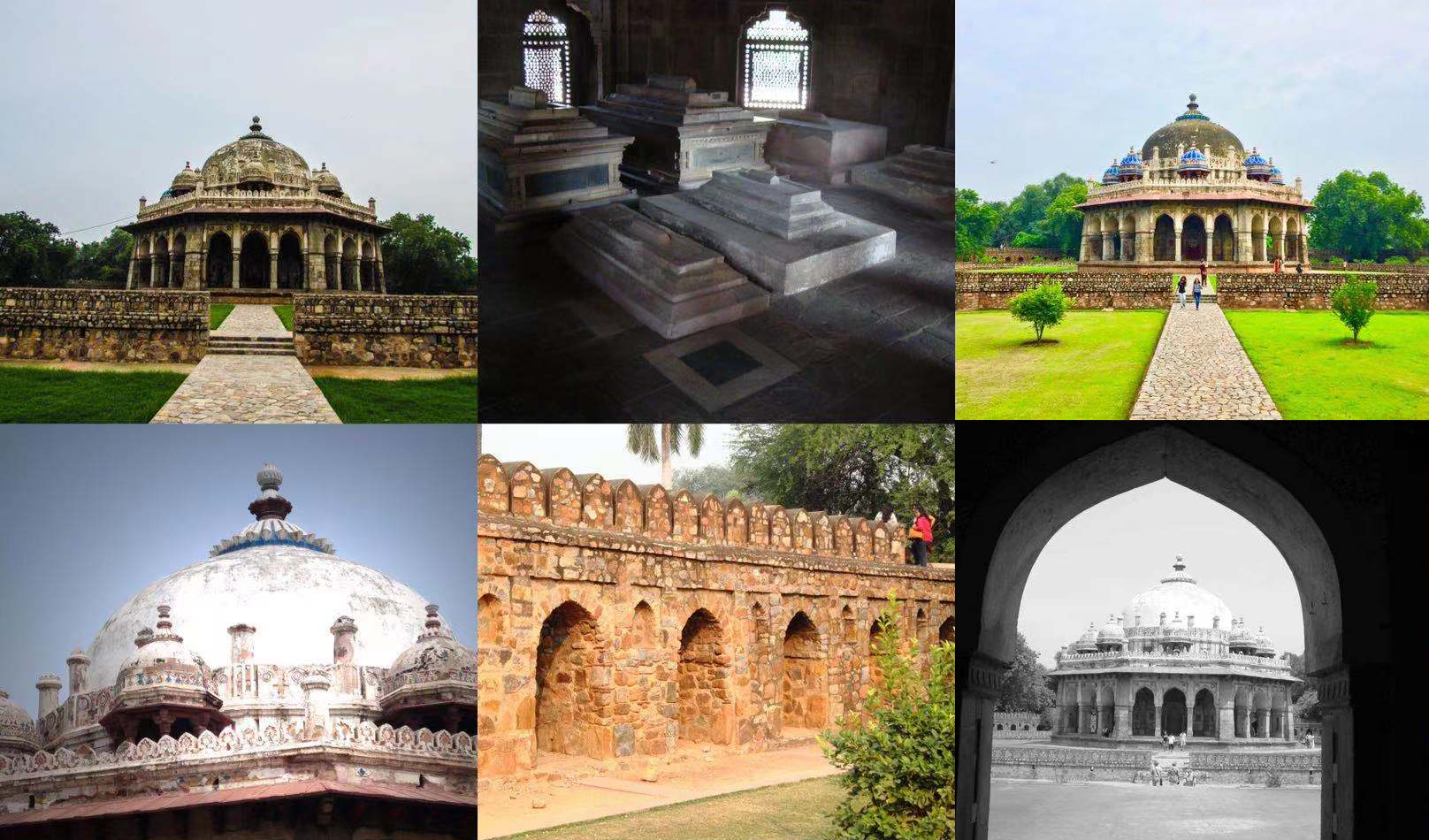}
\caption{Intra-class variance in GLD v2.1 clean data set.}
\label{Intra class variance in GLD-v2.1 clean data set}
\end{figure}

Large intra-class variance is another typical character of landmark datasets: different shooting angles, illumination, and obstacles in the image will cause variation in the image features (see example Fig.~\ref{Intra class variance in GLD-v2.1 clean data set}). 
Recognizing the landmark under such significant noise and interference is quite challenging. \\

In order to reduce the impact of the above-mentioned intra-class variance on the retrieval task, we augment the data (see Table~\ref{Intra-class variance name and data argumentation method in GLD-v2.1 data set.}). 
However, augmenting the data in this way is not sufficient for the Landmark Dataset because it has poor adaptability. 
Therefore, in this paper, we build a \textit{Global-Local} approach to overcome the above challenges. 
The details of our design are demonstrated in the next section. 

\setlength{\tabcolsep}{4pt}
\begin{table}
\begin{center}
\caption{Intra-class variance name and data argumentation method.}
\label{Intra-class variance name and data argumentation method in GLD-v2.1 data set.}
\begin{tabular}{lll}
\hline\noalign{\smallskip}
\textbf{Intra-class variance name} & \textbf{Data augmentation method}\\
\noalign{\smallskip}
\hline
\noalign{\smallskip}
Indoor and outdoor
images & Divide the images into
two categories\\
Season and time period & Color transformation,
image blur
\\
Landmark exterior
maintenance & Random erasing\\
Shooting angles & Random perspective\\
Long shooting and short
shooting & Random crop\\
Blocked view & Random erasing\\
Black and white images & Color transformation\\
Image flip & Random rotation\\
\hline
\end{tabular}
\end{center}
\end{table}
\setlength{\tabcolsep}{1.4pt}
\section{Methods}
\label{Methods}
This chapter is divided into three sections. The first is the overview of the network structure. Sections two and three introduce the global feature extractor and the training process, respectively. The final section introduces the local feature extractor, followed by the retrieval process. 

\subsection{Overview}
\begin{figure}[h]
\centering
\includegraphics[height=0.4\linewidth]{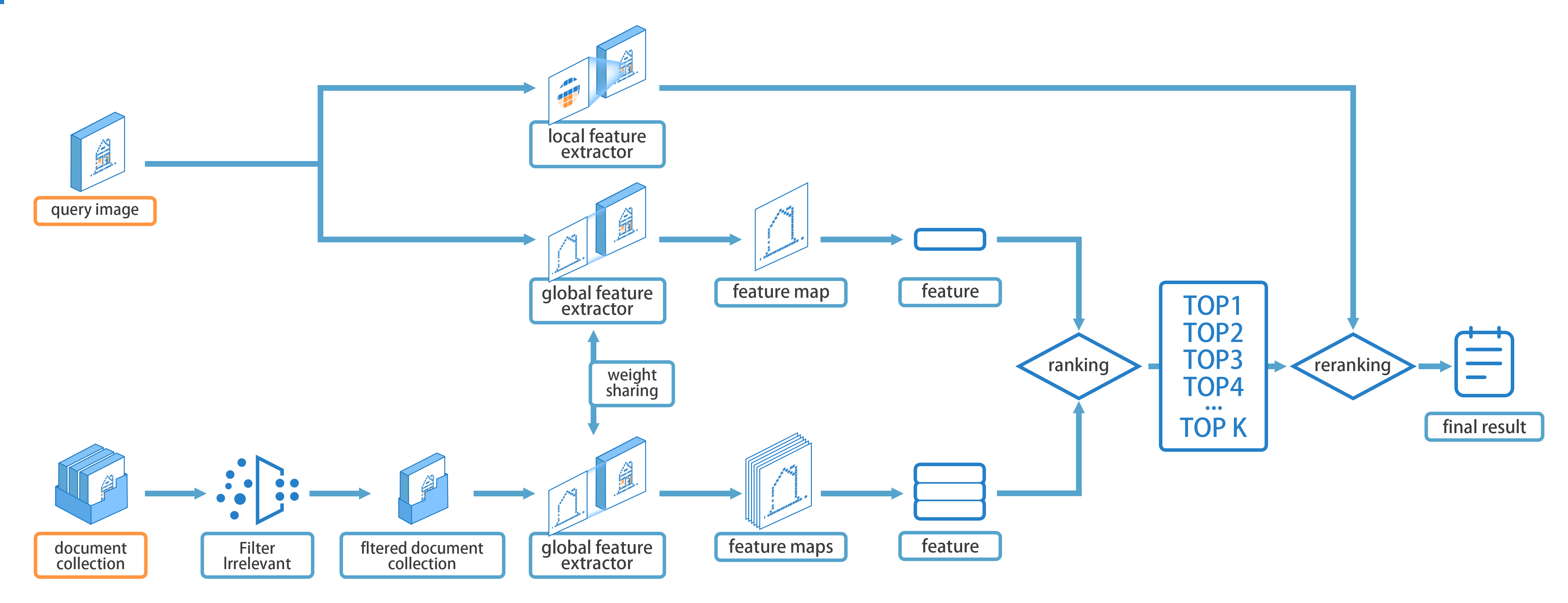}
\caption{The framework of proposed Siamese neural network combining global and local features.}
\label{fig1:framework}
\end{figure}

The network structure in this paper is based on the Siamese network. 
As shown in Fig.~\ref{fig1:framework}, the Siamese network extracts the global features from the query images and the document images respectively. 
Then the document images are ranked according to the similarity of their global features compared with the query image. 
A pre-trained EfficientNet-B0 is employed and fine-tuned, and the cosine similarity of output features is computed, in which the contrastive loss function is applied for optimization. 
This gives the top $K$ similarity images from the document data set. 
This is followed by a re-ranking step based on the local features extractor. 
The top $K$ ($K=100$ in our method) images are re-ranked by comparing local features to the query images. 
Based on the local feature extractor, our global and local feature fusion applies an additional neural network to automatically find the optimal weights for achieving the best combination. 

\subsection{Global feature extractor}
\label{Global feature extractor}
The global feature is a compact representation of the whole image content. 
A commonly used approach to extract the global feature is the Convolutional Neural Network (CNN) \cite{zeiler2014visualizing}. 
In this paper, we adopt the EfficientNet-B0 network, which deeply optimizes the ratio of $D$ (Depth), $W$ (Width), and $R$ (Resolution) of the CNN to achieve a high accuracy with limited network size \cite{tan2019efficientnet}. 
Compared with traditional networks such as VGG \cite{simonyan2014very} and ResNet \cite{he2016deep}, the EfficientNet-B0 has been greatly improved in terms of retrieval speed and accuracy, especially for transfer learning tasks. \\

The feature extracted by CNNs generally has a large number of channels but relatively small feature maps. 
In our work, we use an input image with a size of $224 \times 224 \times 3$, the output feature layer of EfficientNet-B0 is $7 \times 7 \times 1280$. 
In order to reduce the impact of region shift of the image and 
aggregate the feature in a global manner, we apply an average pooling layer to each feature map. 
The average pooling layer treats the contribution of each feature equally and consequently results in an one dimensional feature vector. 
\subsection{Training the global feature extractor}
First, we use transfer learning to fine-tune the pre-trained model (EfficientNet-B0 on the ImageNet data set \cite{5206848}), and the classification layer of the pre-trained model is removed. 
Then a new classification layer is added to fit the network to the
GLD v2.1 data set, which is a fully connected layer with all the categories in the GLD v2.1 data set. 
To speed up the training phase, all the other weights are then frozen and only the last feature layer is trained with the new classification layer. 
After training, the newly added classification layer is removed to obtain a new feature extractor suitable for GLD v2.1 data set. \\

Secondly, after transfer learning the Siamese network and the contrastive loss method are used to fine-tune the new feature extractor. 
Since the input of the Siamese network is an image pair, the process of data mining is as follows: 
to maximize the use of the data set, a batch-wise positive/negative mining process is used to build up all possible positive/negative training pairs within one batch. 
in each batch, the batch sampler divides an even number of images into each category. 
for each image in the batch, 
a miner is used to build up training pairs, which randomly selects one image from the same category as the positive sample or one image from another category as the negative sample. 
finally, the mining process is repeated to build all positive/negative pairs. 
In this way, we can combine the feature extractor of transfer learning and metric learning to fine-tune the network to learn the features of GLD v2.1 data set. 

\subsection{Local feature extractor}
\label{Local feature extractor}
\begin{figure}[H]
\centering
\includegraphics[height=0.65\linewidth]{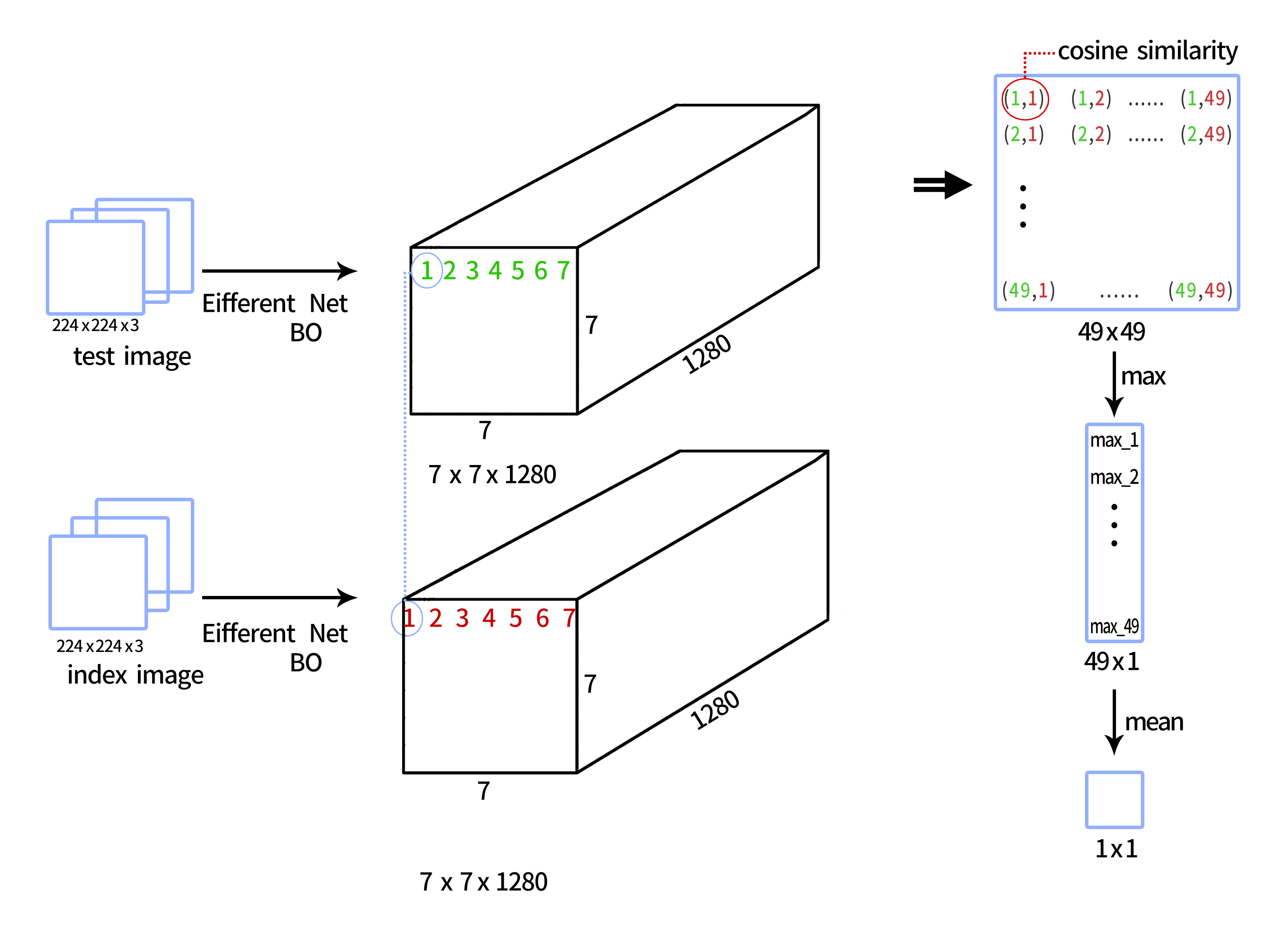}
\caption{The local feature extractor in our method.}
\label{The local feature extractor in our method}
\end{figure}

The local feature focuses on a point or region of interest in the image and represents the contents of a specific image region. 
For extracting the local feature, our method is based on the idea of patch retrieval. 
The local features are incorporated into the initial ranking result from the global feature extractor and the initial ranking result is re-ranked based on the fusion of global and local features. 
Specifically, the local feature extractor is based on the output of neural networks (Fig.~\ref{The local feature extractor in our method}). 
\vspace{-0.75cm} 

First, the global features extracted from EfficientNet-B0 of the test image and the index image are employed. 
As introduced in \ref{Global feature extractor}, the dimension of the obtained feature is  $7\times7\times1280$-dimensional features. 
Second, the global $7\times7\times1280$ features are segmented into 49 local feature patches of 1280 dimensions. 
Intuitively, each patch only describes the local feature of a specific area of the original image. 
Then cosine similarity is used to compare the first local feature patch obtained by the test image segmentation to the 49 local features of the index images each by each. 
Among the 49 comparisons, the highest similarity is regarded as the score of the first local feature. 
Similarly, the rest of the local features of the test image are compared to the index image, producing a total of 49 scores. 
Next, the 49 scores are averaged to get the final local feature score of the two images. 
Then, the global feature score and the local feature score are fused by neutral network, which is also trained on GLD v2.1, to get the final score of similarity. 


\subsection{Retrieval process}
\label{Retrieval process}
In this subsection, we introduce the retrieval process that follows the training process. 
Before the retrieval process starts, the global feature extractor is used to extract the global features of the index dataset embedding. 
For a given landmark image, the target of the retrieval is to find images with same landmark from the index data set. 
First, the landmark image is embedded in the EfficientNet-B0 feature extractor to obtain the global feature. 
Next, the cosine similarity algorithm is used to calculate the similarity of the query image to the images in the index data set, and the 100 images with the highest similarity are output in descending order and re-ranked using local features. 
In the second stage, we conduct the re-ranking process on the results of the initial ranking, using the local feature extractor. 
The local features from these images are extracted and feature patch matching is performed with the query images. 
Finally, the performance of EfficientNet-B0 in the test data set and index data set is evaluated by calculating the resulting mAP$@$100. 

\section{Experiment and Analysis}\label{Experiment}
This chapter discusses the specific methods by which the evaluation of the model is conducted. 
This chapter is divided into three subsections, which demonstrate the outstanding performance of the proposed method. 

\subsection{Settings}
\subsubsection*{Training Settings}
The network is trained on GLD v2.1 data set containing 1.5 million landmark images for 20 epochs. 
The Adam optimizer is used with an initial learning rate of $1.5\times 10^{-4}$, and the Adam parameter is selected to be $10^{-8}$. 
\subsubsection*{Evaluation Settings}
The model is evaluated on a test data set containing 1000 landmark images, which is sampled from GLD v2.1 test data set in such a way that each test image label is guaranteed to appear at least once in the training data set. 
For evaluation purposes, the test data set is split into a private set (650 images) and a public set (350 images). 
As for the target of retrieval, the index data set contains 100 thousand images in total. 
Finally, regarding the ranking result, the mAP@100 score is calculated. 
For comparison experiments, we adopted ResNet-18, ResNet-50 and EfficientNet-B0 for the global feature extractor and Scale-Invariant Feature Transform (SIFT)\cite{lowe2004distinctive}, and Vector of Locally Aggregated Descriptors (VLAD) \cite{jegou2010aggregating} for the local feature extractor. 
The results are compared and discussed. 
\subsection{Experimental result analysis}
The landmark retrieval results of other traditional networks have been completed and the experimental results are shown in \autoref{Retrieval result mAP@100.}. 
ResNet-18 and ResNet-50 are used to replace the EfficientNet-B0 network in the applied method. 
It can be seen that when ResNet-18 is used, the accuracy is only 16.96\% (private) / 17.30\% (public). 
When ResNet-50 is used, the effectiveness of the extracted features increases due to the increase in the depth of the network, and the retrieval accuracy reaches 22.76\% (private) / 23.41\% (public), but this is still far below the accuracy of the method proposed in this work. 
When using EfficientNet-B0 to fine-tune only the last layer, the result of transfer learning is 25.43\% (private) / 25.59\% (public). 
This is because the method cannot effectively learn inter-class variance, since it doesn't learn pairwise similarities, as a Siamese Network does. 
The method proposed in this work can simultaneously use EfficientNet for efficient feature extraction, learn inter-class variance with the help of Siamese network structure, and use the re-ranking method based on local features to find subtle feature changes and further improve retrieval accuracy. 
\setlength{\tabcolsep}{4pt}
\begin{table}
\begin{center}
\caption{Retrieval result mAP@100.}
\label{Retrieval result mAP@100.}
\begin{tabular}{lll}
\hline\noalign{\smallskip}
\textbf{Method} & \textbf{Private} & \textbf{Public} \\
\noalign{\smallskip}
\hline
\noalign{\smallskip}
Our method & \textbf{33.09\%} & \textbf{32.84\%} \\
ResNet-18 & 16.96\% & 17.30\%\\
ResNet-50 & 22.76\% & 23.41\% \\
EfficientNet-B0 & 25.43\% & 25.59\% \\
\hline
\end{tabular}
\end{center}
\end{table}
\setlength{\tabcolsep}{1.4pt}

\subsection{Ablation study}
In this subsection, first we will compare the effect of different fine-tuning methods on global features, then different local features will be compared.

The baseline uses pre-trained EfficientNet-B0 as a feature extractor. 
The purpose of this is to test the behavior of the original EfficientNet-B0 on the landmark retrieval task and verify the improvement of transfer learning in the following experiments. 
The fine-tuning based on transfer learning is used to transfer the weight of EfficientNet-B0 from the ImageNet data set to the GLD v2.1 data set. 
The classification layer of the pre-trained model is replaced according to GLD v2.1 data set, and the last feature layer is trained by GLD v2.1 data set. 
The metric learning-based fine-tuning is used by adopting the Siamese network. 
The feature extractor follows the structure of baseline. 
The initial ranking result of our proposed method is also listed to compare the performance of different global features. 

\begin{figure}[H]
	\centering
	\subfloat[Baseline.]{
		\begin{minipage}[b]{0.45\textwidth}
			\includegraphics[width=\textwidth]{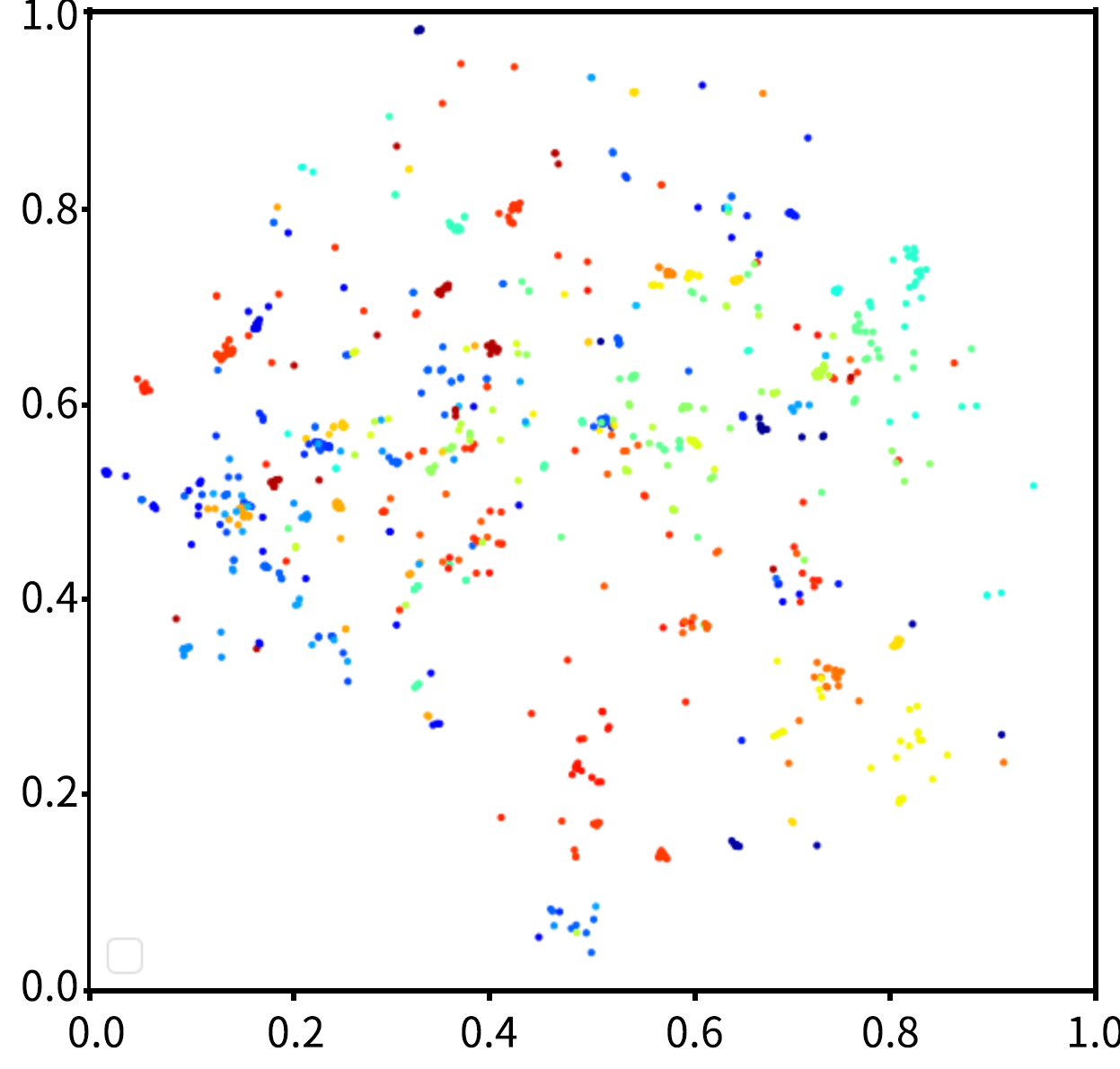} 
		\end{minipage}
		\label{Experiment 1 randomly selected 50 categories}
	}
    	\subfloat[Transfer learning-based fine-tuning.]{
    		\begin{minipage}[b]{0.45\textwidth}
  		 	\includegraphics[width=\textwidth]{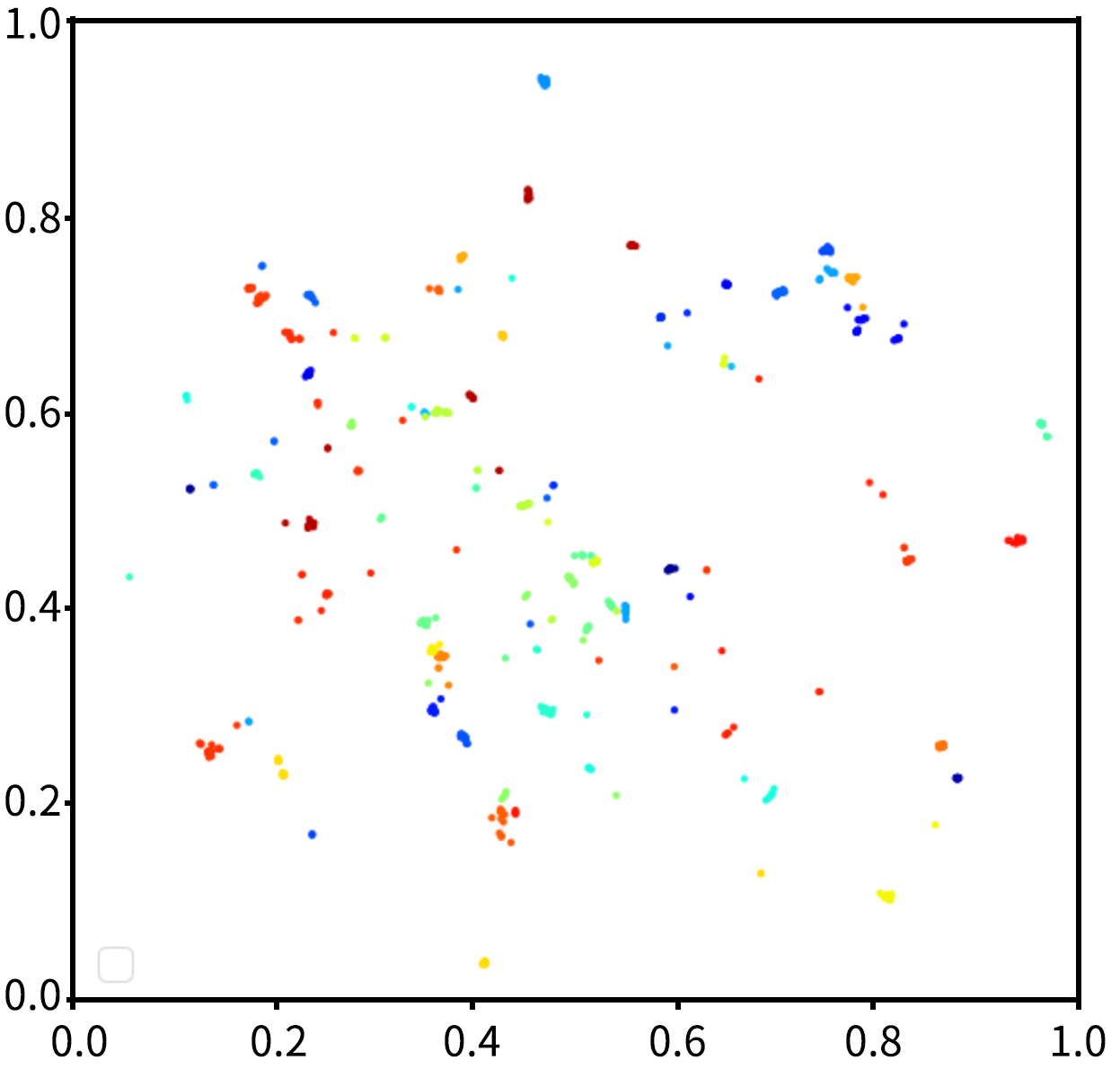}
    		\end{minipage}
		\label{Experiment 2 randomly selected 50 categories}
    	}
	\\ 
	\subfloat[Metric learning-based fine-tuning.]{
		\begin{minipage}[b]{0.45\textwidth}
			\includegraphics[width=\textwidth]{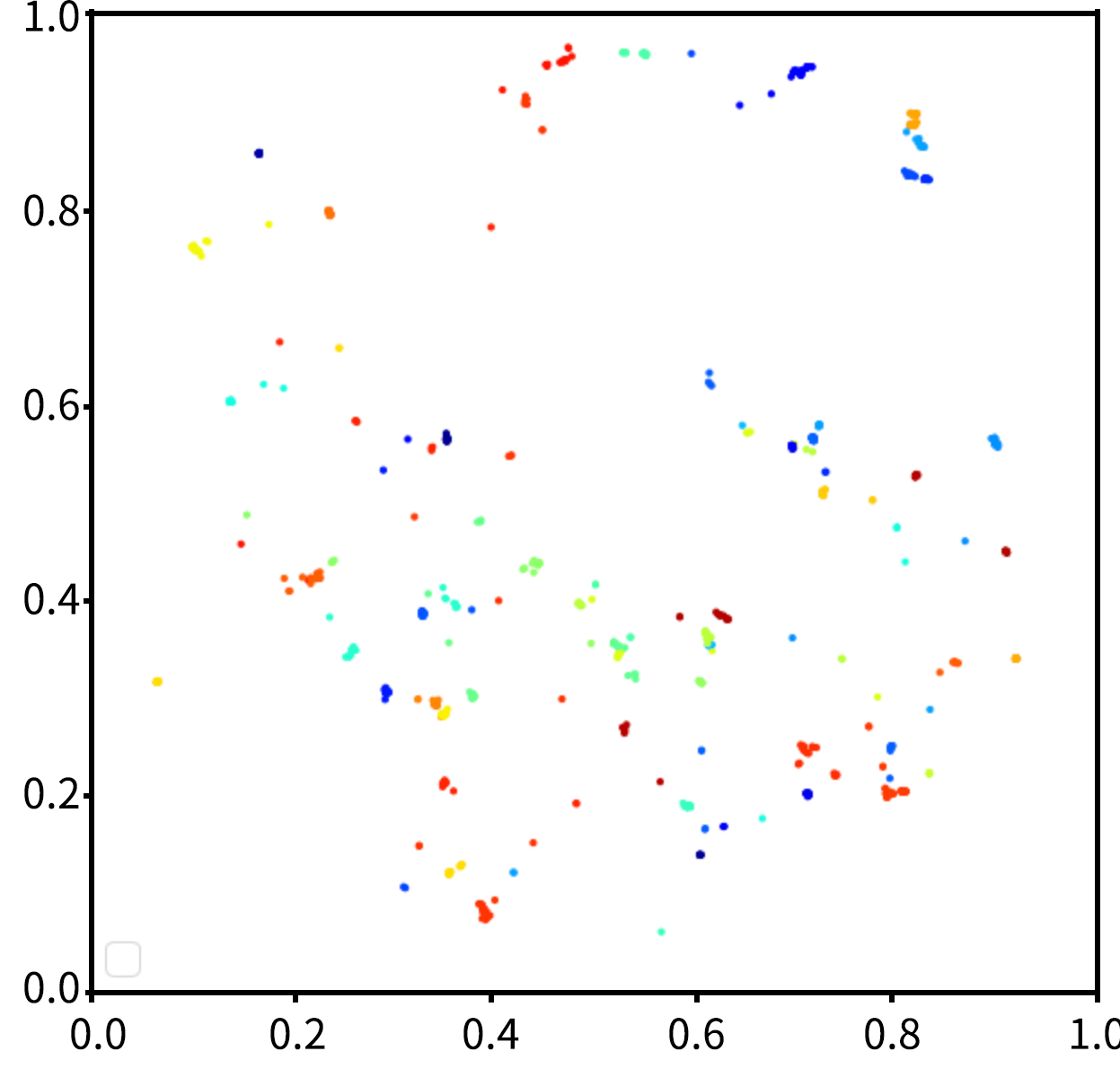} 
		\end{minipage}
		\label{Experiment 3 randomly selected 50 categories}
	}
    	\subfloat[Initial ranking result.]{
    		\begin{minipage}[b]{0.45\textwidth}
		 	\includegraphics[width=\textwidth]{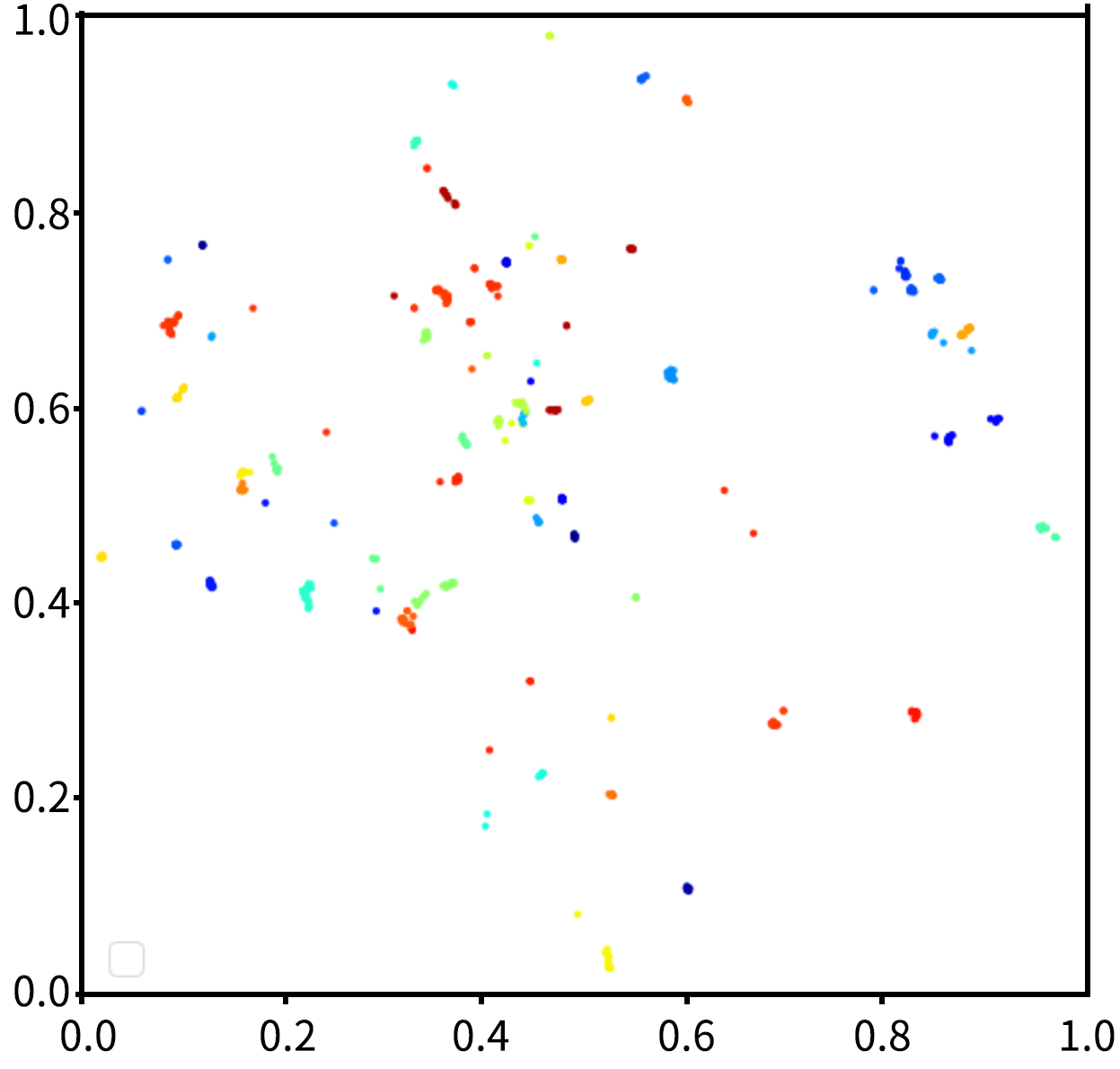}
    		\end{minipage}
		\label{Experiment 4 randomly selected 50 categories}
    	}
	\caption{t-SNE result of 50 randomly selected categories.}
	\label{t-SNE result randomly selected 50 categories}
\end{figure}

Fig.~\ref{t-SNE result randomly selected 50 categories} shows an example of global feature projection using t-distributed Stochastic Neighbor Embedding (t-SNE) \cite{van2008visualizing} with 50 randomly selected categories. 
Because of the large number of categories, Fig.~\ref{t-SNE result randomly selected 50 categories} may show similar colors for different categories. 
As can be seen from Fig.~\ref{t-SNE result randomly selected 50 categories}, going from (a) to (d), the intra-class variance of the sample is reduced, and the inter-class variance is increased. That is to say, the similarity within the category increases, and the difference between the categories also increases. \\As a result, the global feature extractor of our proposed method is the best overall among the fine-tuning methods assessed. 
\vspace{-0.8cm}
\vspace{-0.3cm}
 
Next, to compare the re-ranking result, SIFT, VLAD and our local feature extractor are respectively implemented based on the initial ranking result. 
All results are listed in  Table~\ref{The mAP@100 result of local feature re-ranking.}. \\
\vspace{-0.5cm}
\begin{table}
\begin{center}
\caption{The mAP@100 result of global and local feature re-ranking.}
\label{The mAP@100 result of local feature re-ranking.}
\begin{tabular}{lll}
\hline\noalign{\smallskip}
\textbf{Method} & \textbf{Private} & \textbf{Public}\\
\noalign{\smallskip}
\hline
\noalign{\smallskip}
Baseline  & 8.23\% & 7.42\% \\
Transfer learning-based fine-tuning & 25.43\% & 25.59\% \\
Metric learning-based fine-tuning & 31.49\% & 31.68\% \\
Ours (initial ranking result)  & \textbf{32.57\%} & \textbf{32.68\%}\\
\hline
SIFT (re-ranking result) & 32.61\% & 32.71\% \\
VLAD (re-ranking result) & 30.91\% & 30.85\% \\
Ours (re-ranking result) & \textbf{33.09\%} & \textbf{32.84\%} \\
\hline
\end{tabular}
\end{center}
\end{table}
\setlength{\tabcolsep}{1.4pt}

The result of baseline is 8.23\% (private) / 7.42\% (public). 
This baseline performance is only 8\% because the EfficientNet-B0 model was initially trained on the ImageNet data set.  
The categories contained in the GLD v2.1 data set are basically human-made buildings. 
Hence, most of the categories in ImageNet do not exist in GLD v2.1. 
As a result, the original network is trained to expand the distance between different items in order to classify images of different types of items. 
In contrast, GLD v2.1 data sets are all architectural images and their category scope is much smaller than ImageNet. Fine-tuning must be carried out to expand the intra-class differences. \\

The result of fine-tuning based on transfer learning is 25.43\% (private) / 25.59\% (public). 
It can be observed from mAP@100 that compared to baseline, the fine-tuned model improves retrieval accuracy considerably. 
Note that this method has fine-tuned the feature extractor of baseline, which makes the retrieval results more suitable for the GLD v2.1 data set. 
As a result, a landmark image will extract less information about horses, cars, buckles, and other items but pay more attention to landmark-related information. \\

The result of metric learning-based fine-tuning is  31.49\% (private)/ 31.68\% (public). 
In this method, the structure of the Siamese network is added based on baseline. 
Due to the conversion of samples to sample pairs, the method has increased the sample size compared with transfer learning, thus reducing the problem of overfitting. 
The method can distinguish the inter-class distance compared to the baseline, but since it only utilizes the EfficientNet pre-trained features on ImageNet, the method cannot extract accurate landmark features. 
Therefore, the accuracy is still unsatisfactory. \\

\begin{figure}
\centering
\subfloat[Baseline.]{
	\begin{minipage}[t]{\linewidth}
	\includegraphics[width=\linewidth]{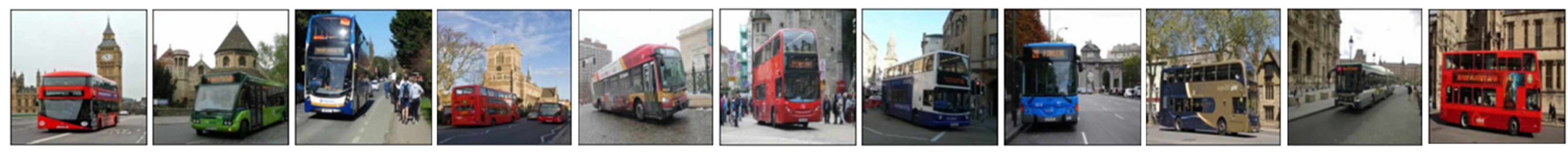} 
    \end{minipage}
	}\\
\subfloat[Transfer learning-based fine-tuning.]{
	\begin{minipage}[t]{\linewidth}
	\includegraphics[width=\linewidth]{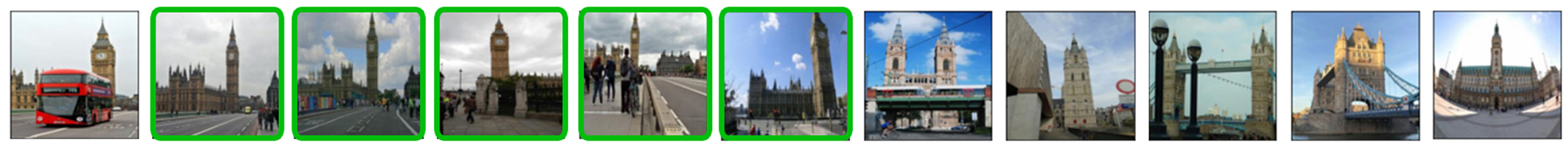} 
	\end{minipage}
}
\\
\subfloat[Metric learning-based fine-tuning.]{
	\begin{minipage}[t]{\linewidth}
	\includegraphics[width=\linewidth]{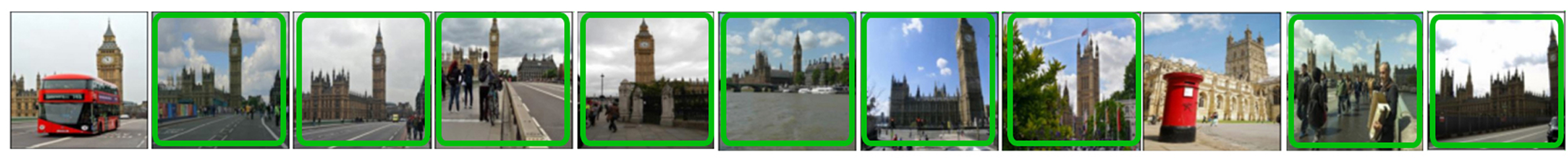} 
	\end{minipage}
}
\\
\subfloat[Initial ranking result.]{
	\begin{minipage}[t]{\linewidth}
	\includegraphics[width=\linewidth]{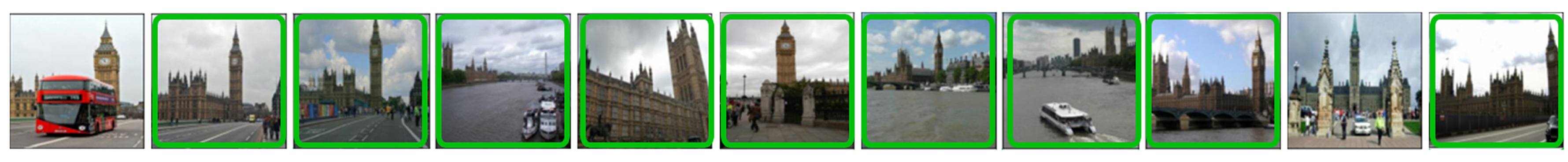} 
	\end{minipage}
}
\caption{Result of all fine-tuning methods of global feature (query image on the left side, corrective retrieval results marked in green box).}
\label{Result of Experiment 1 to Experiment 4}
\end{figure}

The initial ranking result of our method is  32.57\% (private)/ 32.68\% (public). 
Compared with transfer learning, our method uses the pair mining algorithm to enlarge the sample size, which reduces the over-fitting problem and enhances the learning of the features between intra-class. 
Compared with metric learning, our method uses the outcome of transfer learning as a feature extractor, which is more suitable for GLD v2.1 information extraction than only using metric learning. 
Therefore, our method simultaneously fine-tunes the original EfficientNet and trains the network with the Siamese network, which is the most suitable structure for training the GLD v2.1 data set in the global features. 
The retrieval results of (a) to (d) are shown in Fig.~\ref{Result of Experiment 1 to Experiment 4}, with the query image on the left. We output the top 10 retrieval results and we can see that (d) is the most accurate. \\ 

The Table~\ref{The mAP@100 result of local feature re-ranking.} shows the re-ranking results of our initial ranking result using different local features. 
It can be seen that both SIFT and our methods further improve the accuracy of retrieval. Specifically, SIFT slightly improves the search accuracy by 0.04\% (private) / 0.03\% (public). 
The $Global + Local$ approach has the best results, increasing 0.48\% (private) / 0.16\% (public). 
This shows that the local feature extractor based on the global feature extracted by EfficientNet-B0 can extract local features more accurately than the hand-designed SIFT operator. 
For the VLAD feature, the retrieval accuracy has declined. 
The reason is that only the cluster center $k = 256$ is set for the codebook training, due to the computational limit.
This is too small for the GLD v2.1 data set with 81,024 categories, which may lead to the model under-fitting. 
The SIFT and VLAD local feature extraction methods both require a large amount of calculation time because of the inclusion of the clustering process. 
In contrast, our patch retrieval method is simpler and more effective. \\

In summary, it has been shown from experiments that re-ranking global and local features using the EfficientNet-B0 and Siamese network is an improvement on traditional networks. 
The best mAP@100 is achieved when we use the Siamese network and our patch retrieval, and use the neural network in the re-ranking process to get the best results.

\section{Conclusions}
In this paper, we build a two-stage network $Global-Local$, consisting of global and local feature extractors, in which a Siamese network extracts global features from the query images and the document images respectively. 
Moreover, we use patch retrieval methods to optimize retrieval accuracy of global features. We evaluate our model on GLD v2.1 dataset. Experimental results show that our model outperforms the state-of-the-art method (ResNet), demonstrating the wide potential of applying our model to real-world applications. The cosine similarity followed by pooling provides a simple method to refine image results. Furthermore, we analyzed the Google Landmark Dataset and applied several data augmentations to deal with various intra-class variances. \\ 

In future work, we would like to test our approach on other architectures and improve the patch retrieval process using an attention mechanism. Although the cosine similarity enables the comparison of image patches in the embedding space, it is not a trainable part of the architecture. It could be useful to integrate a cross-attention mechanism that allows the comparison of image patches in a differentiable way.



\clearpage
%
%
\bibliographystyle{splncs04}
\bibliography{egbib}
\end{document}